\documentclass[english,a4paper,oneside,twocolumn,aps]{revtex4}
\usepackage[english]{babel}
\usepackage[utf8x]{inputenc}
\usepackage{microtype}
\usepackage{graphicx}
\usepackage{amsmath}
\usepackage{siunitx}
\usepackage[colorlinks=true,linkcolor=black,citecolor=black]{hyperref}
\usepackage{epstopdf}
\usepackage[capitalise]{cleveref}

\date{\today}
\begin{document}

\title{Window for Efimov physics for few-body systems with finite-range interactions}
\author{S. E. Rasmussen\footnote{E-mail: stig.elkjaer.rasmussen@post.au.dk}, A. S. Jensen, and D. V. Fedorov}
\affiliation{Department of Physics and Astronomy, Aarhus University, DK-8000 Aarhus C, Denmark}

\begin{abstract}
We investigate the two lowest-lying weakly bound states of $N \leq 8$
bosons as functions of the strength of two-body Gaussian interactions.
We observe the limit for validity of Efimov physics.  We calculate
energies and second radial moments as functions of scattering length.
For identical bosons we find that two $(N-1)$-body states appear
before the $N$-body ground states become bound.  This pattern ceases
to exist for $N \geq 7$ where the size of the ground state becomes
smaller than the range of the two-body potential.  All
mean-square-radii for $N \geq 4$ remain finite at the threshold of
zero binding, where they vary as $(N-1)^{p}$ with $p=-3/2,-3$ for
ground and excited states, respectively.  Decreasing the mass of one
particle we find stronger binding and smaller radii.  The identical
particles form a symmetric system, while the lighter particle is
further away in the ground states.  In the excited states we find the
identical bosons either surrounded or surrounding the light particle
for few or many bosons, respectively.  We demonstrate that the first
excited states for all strengths resemble two-body halos of one
particle weakly bound to a dense $N$-body system for $N=3,4$.  This
structure ceases to exist for $N \geq 5$.
\end{abstract}

\maketitle

\section{Introduction}\label{sec:intro}	

The original Efimov effect \cite{EfimovPhysLett1970,EfimovSoviet1971} describes the three-body structures using a single two-body parameter, the scattering length. Three identical bosons form infinitely many bound states with an accumulation point at the two-body scattering threshold where the scattering length approach infinity. 
The Efimov effect has universality and scale invariance as important concepts, and from generalizing these concept emerges Efimov physics which we in this article understand as the quantum physics where universality and scale independence apply. 

We define universality as independence upon the shape of the interparticle potential, hence any interparticle model-potential should yield the same results. Scale invariance means independence upon the length scale of the system, hence the Efimov physics should be applicable on nuclei, atomic or any other scale.
These concepts are unfortunately often only implicitly defined and with the meaning exchanged.
Closely related to the Efimov effect is Borromean binding. This refers to a bound three-body system where none of the subsystems is bound. Thus removing just one of the bodies in such a system would make it unbound. This is often pictured classically by the three Borromean rings. Borromean systems often have large spatial extension and are weakly bound. Thus their structure is determined solely by the binding energy, which makes them good candidates for universality.

Efimov originally suggested that the effect could be found in nuclei, however the evidence was first reported for an ultra cold gas of caesium atoms using their three-body recombination rates \cite{Kraemer2006}. It has since been observed in other ultracold gases of alkali metals (potassium and lithium)  by exploiting Feshbach resonances to vary the effective two-body interaction \cite{Zaccanti2009,Gross2009,Pollack2009}, and more recently also in the Helium trimer \cite{Kunitski2015}.

Following the realization of the Efimov effect, much effort has been put into Efimov physics in general \cite{DIncao2017,Greene2017,Naidon2017}, expanding it beyond three identical bosons. These concepts are presented in the following two subsections.

\subsection{Efimov physics beyond three bosons}

Three years after Efimov predicted the effect, it was proven that there is no true Efimov effect for $N\geq 4$ \cite{Amado1973}. It was later predicted that each Efimov state is accompanied by two universal Borromean tetramer ($N=4$) states \cite{Hammer2007,Stecher2009}. This has been confirmed experimentally \cite{Pollack2009,Ferlaino2009}, which therefore raises the question, whether this pattern continues for higher $N$. Theoretical calculations for $N = 5,6$ have been made, which predict that the pattern of two $N+1$ state accompany each universal $N$ ground state at least for $N\leq 6$  \cite{Brodsky2006,Thogersen2008,Deltuva2010,Stecher2010,Stecher2011,Deltuva2012,Gattobigio2012,Bazak2016,Horinouchi2016,Kievsky2017}. Experimental evidence that two Borromean pentamer ($N=5$) states does accompany the ground tetramer state have been reported \cite{Zenesini2013}. 

Many calculations exploit the simplicity of zero-range interactions with two well-known principle problems, that is, the necessary regularization due to the small distance divergence and the missing finite range of a realistic interaction.  The latter short-coming is essential in investigations touching upon or searching for the limits of Efimov physics.  Quantum states with a substantial part located inside the range of the potential between neighboring particles must depend on the characteristics of the potential and consequently they cannot be universal.  It is therefore essential to employ a finite-range potential for such investigations.  Once the range is introduced in the practical calculations, the radial shape is unimportant for investigations of Efimov physics when the range of the potential is much smaller than the scattering length of the system.  We shall use a Gaussian shape throughout this paper.

In \cref{sec:Iden} of this paper we present results for ground and first excited states for systems of $N\leq 8$ identical bosons. This is done via numerical calculations using correlated Gaussian expansion determined by stochastic variation which is briefly described in \cref{sec:svm}.  We vary binding energies from zero to well-bound to investigate the limits for validity of universality.

The structure of $N$-body systems is usually examined by calculating the pair-distribution function \cite{Stecher2010,Stecher2011,Blume2014}, however in this paper we present the actual radius of the system, and the average distance between the nearest neighboring particles for the ground and first excited state. For universality to apply we would require these to remain larger than the range of the potential. As $N$ increases these distances become shorter and hence we would expect that universality is broken at some $N$. 

For illustrative purposes we compare in \cref{sec:schem} our numerical calculations with schematic models. We compare with analytic harmonic oscillator results and estimates from Gaussian approximations of the true density distributions.

\subsection{Efimov physics beyond identical bosons}

The Efimov effect does not only occur for three identical bosons, but also for other three-body systems, with large scattering lengths. Efimov himself considered particles of different mass, and showed that the effect occurs for any mass ratio between the three particles \cite{Efimov1973}. This has been verified experimentally using alkali atoms of different mass, rubidium and potassium in \cite{Barontini2009} and caesium and lithium in \cite{Pires2014,Tung2014}.  The scaling between the universal states of the system only depends on the mass ratio of the particles, making this quantity the parameter of interest when investigating non-identical particle systems.  Mass asymmetric systems have been studied rather extensively both experimentally, \cite{Barontini2009,Bloom2013,Pires2014,Tung2014,Maier2015,Ulmanis2015,Ulmanis2016,Wacker2016,Johansen2016} and theoretically \cite{DIncao2006,Helfrich2010,Wang2012b,Mikkelsen2015,Petrov2015,Fedorov2015}. 
Lately it has even been discovered that a so-called {\it super}-Efimov effect among fermions in two dimensions \cite{Levinsen2008,Nishida2013,Volosniev2014,Moroz2014,Efremov2014,Gridnev2014,Gao2015}, and also Efimov scaling phenomena in the dynamics of a strongly-interacting many-body Fermi gas \cite{Deng2016}.

Theoretical calculations for more than three bosons of different mass provide energy scalings between different states \cite{Blume2014,Schmickler2017}, while leaving the spatial configuration unpredicted.  In \cref{sec:mass1} we thus investigate the three- to seven-body systems where the mass of one particle is varied, with focus on resolving the spatial configuration between the particles. We numerically calculate the ground and first excited states of these systems, where we vary the mass of one of the particles until its mass is a twentieth of the rest of the particles.  We investigate the systems at the unitarity limit (infinite two-body scattering length) for scattering between unequal pairs of particles, while also maintaining a large scattering length for pairs of identical particles.

According to Efimov there should also be Efimov states for one heavy particle and two lighter particles.  However, such Efimov states would be very loosely bound and are hence suppressed in low energy Efimov experiments \cite{Tung2014}.  Efimov systems with one heavy particle are not only experimentally difficult to produce, they are also difficult to calculate numerically again due to the weak binding.  The same is true for more than three particles.  Hence we focus on systems where one of the particles has less mass than all the other $N-1$ particles.

\section{Formulation of problem}\label{sec:formu}

For the first part of this paper we consider $N$ identical bosons, of mass $m$, with coordinates $\mathbf{r}_i$ and moment $\mathbf{p}_i$. We restrict the particles to interact only via two-body short-range potentials. Hence the Hamiltonian becomes
\begin{align}
H = \sum_{i=1}^{N} \frac{\mathbf{p}^2_i}{2m} - T_\text{CM} + \sum_{i<j}^{N} V(r_{ij})\, ,
\end{align}
where $r_{ij} = |\mathbf{r}_i - \mathbf{r}_j|$ and $T_\text{CM}$ is
the kinetic energy of the uninteresting decoupled center-of-mass motion.

In the second part of this paper where the mass of one of the particles, $M$ is varied, the Hamiltonian becomes
\begin{equation}
\begin{aligned}
H =& \sum_{i=1}^{N-1} \frac{\mathbf{p}^2_i}{2m} + \frac{\mathbf{p}^2_N}{2M}  - T_\text{CM}\\
 &+ \sum_{i<j}^{N-1} V_{HH}(r_{ij}) + \sum_{i=1}^N V_{HL}(r_{iN})\, ,
\end{aligned}
\end{equation}
where $V_{HH}$ is the interaction between the identical (heavy-heavy) particles, and $V_{HL}$ is the interaction between interspecies (heavy-light) particles. For the present calculations we will however assume that $V_{HH}=V_{HL}$ and therefor omit the subscript.

As we are interested in universality we can in principle choose any finite-range two-body potential. For simplicity we choose a Gaussian potential, such that the interaction between the $i$th and $j$th particle becomes
\begin{align}\label{eq:guass}
V(r_{ij}) = V_0e^{-r_{ij}^2/b^2},
\end{align}
where $V_0$ is the potential strength, which is tuned in order to change the scattering length, $a$, of the two-body system. The range of the potential, $b$, is the characteristic length scale of the interaction, and together with the characteristic energy 
\begin{align}\label{eq:CharE}
E_s = \frac{\hbar^2}{2\mu b^2}\, ,
\end{align}
where $\mu$ is the two-body reduced mass, it defines the universal regime, by requiring that the energies are much smaller $ E_s$ and the length scales much larger than $b$.  Since we consider a finite-range potential, we expect some range-corrections \cite{Thogersen-finite2009,Sorensen2013a} compared to a zero-range potential. We also note that the finite-range potential implies that we do not need to provide a three-body cut-off at short-range of the order the van der Waals length
\cite{Berninger2011,Chin2011,Schmidt2012,Naidon2012,Wang2012a,Sorensen2012,Sorensen2013b,DIncao2013,Roy2013,Naidon2014,Hiyama2014,Horinouchi2015}.

\subsection{The stochastic variational method}\label{sec:svm}

All numerical calculations are done using the stochastically correlated Gaussians as trail functions
\begin{align}
\left|\psi\right\rangle = \exp \left\lbrace -\frac{1}{2}\sum_{j<i}^N \alpha_{ij}r_{ij}^2 \right\rbrace,
\end{align}
where the $N(N-1)/2$ parameters $\alpha_{ij}$ are stochastically chosen. These trial functions are chosen since they are easily generalized to $N$-body problems, and approximate all wave functions well. Furthermore they make an analytical calculation of matrix elements possible. Lastly the functions are spherically symmetric hence the calculated system has the total angular momentum equal zero.

A basis of these trial functions is initially chosen stochastically. The Generalized Ritz Theorem ensures that the expectation value of the Hamiltonian, restricted to the subspace of the the stochastically chosen trial functions, is always above the actual energy.
Testing for a number of different sets of trial functions, the basis which yields the lowest energy is chosen, and the size of the basis is increased by one. Increasing the size of the basis will alway decrease the variational energy, and thus bring it closer to the exact value. Once again different sets of trial functions are tested and the one yielding the lowest energy is chosen, and such the process continues until the expectation value of the energies stabilizes and the calculation is stopped. At this point the basis of trial functions approximate the actual Hilbert space of the system, and thus other quantities of interest can be reliably calculated.

Since we are indeed approximating the Hilbert space we expect some numerical imprecision. These imprecisions are larger for more excited states and closer to the threshold of binding as these become more difficult to calculate. A method for lessening this numerical noise is to increase the basis of trial functions, however as the computational difficulty of this method increases as the square of the number of trial functions, this cannot continue forever. All numerical noise cannot be eliminated, and we allow some imprecision in the results appearing as small fluctuations. \cite{Suzuki1998,Brac2007}

\section{Identical particles}\label{sec:Iden}

\begin{figure*}[t!!!]
	\centering
	\includegraphics[width=\textwidth]{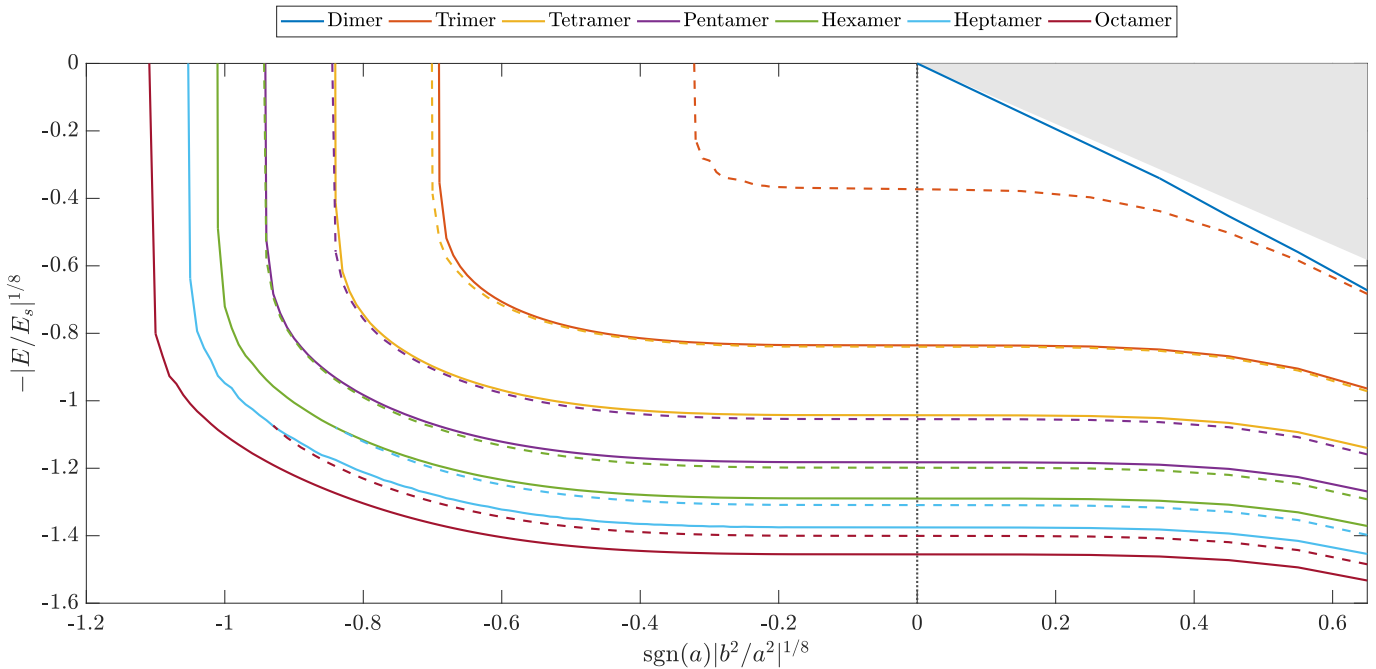}
	\caption{The energy of an $N$-body system consisting of
		identical particles, in units of the
		characteristic energy, \cref{eq:CharE}, as a function of the inverse squared
		scattering length, in units of $1/b$. The figure has been
		scaled to the $1/8$ power in order to increase
		visibility. The solid lines are the ground states, while the
		dashed lines are the first excited states. The dotted
		vertical line indicates the binding threshold for the dimer
		(unitarity line), while the solid blue line in the upper
		right part is the dimer binding energy above which the dimer
		may be unbound. The gray area indicates the unbound region for a zero-range dimer.}
	\label{fig:scat}
\end{figure*}

We consider $N$-body systems of identical particles interacting
pairwise with a Gaussian potential, \cref{eq:guass}.  The particles may
have integer or half-integer spin but we assume the interaction is
spin independent. Thus the spatial wave function should be
supplemented by the spin part and the total symmetry correspondingly
derived. 

We start numerically by using a basis of a few hundred trial
functions.  When we approach the threshold of binding a larger basis
is needed since the systems are very weakly bound and often spatially
extended.  Shifting away from the threshold towards more binding, 
convergence is reached with a smaller basis. In order to further improve the convergence of the calculation the stochastically chosen parameter is steered towards numerically larger values near threshold, as the stochastically parameter is directly related to the spatial extend of the system.

\subsection{Energies}

We calculate the energies and two-body scattering lengths of the
systems for different potential strengths, $V_0$.  We plot the $N$-body
energies as functions of the square of the inverse scattering length
(a measure of the two-body energy) in \cref{fig:scat} for $N \leq 8$.
Following the by now established tradition, the axes in \cref{fig:scat} has been scaled in order to increase visibility.

We first notice that both trimer states lie below the dimer ground state.  We do not calculate more trimer states as only the two first states for all other $N$ are of interest in this paper, but infinitely many trimer states exist there below the dimer continuum as Efimov predicted.  The binding of the excited state increases slower than the dimer binding energy, which is reached at some point after the dimer
has become bound.

In order to assess the effect of the finite-range interaction we calculate the scaling factor defined here as the ratio between square roots of energies between the two lowest trimer states. Between the two lowest states we find this value numerically equal $23.1$ at unitarity i.e. where the dimer binding is zero.
This should be compared to the value for a zero-range interaction found analytically to be
\begin{align}\label{eq:3scalling}
S_3 = e^{\pi/s_0} \simeq 22.7,
\end{align}
where $s_0 \simeq 1.00624$. This is however the scaling factor of $\sqrt{E_3^{(n)}/E_3^{n+1}}$ when $n$ approaches infinity. Since we are only dealing with the two lowest states, we should expect that the numerical scaling factor deviates some what from the analytical result in \cref{eq:3scalling}. The discrepancy in the scaling factors can be attributed to the finite range of the potential. This is due to the fact that even though the scattering length might be infinite, the size of the system is not. In column six and seven of \cref{tab:prop} the average distances between the nearest neighboring particles are presented at the threshold of binding. For the two lowest trimer states these distances 5.9 and 50 times the range of the potential respectively. At the unitary limit these distances are even smaller, as seen on \cref{fig:idenDist}. Thus the finite range is responsible for the discrepancy between these scaling factors. We further compare our results with \cite{Gattobigio2014}, who also calculates this scaling factor with a two-body finite Gaussian interaction, and finds $23.0$, in good agreement with our result. The relation for the higher-lying excited trimer states would asymptotically approach the analytic result.

The behavior is completely different for $N > 3$.  The attraction needed for binding decreases with increasing $N$, and the spectra for each $N$ are limited to finite numbers below the trimer threshold.  It is therefore inevitable that only a finite number of $N$-body states can appear before the trimer becomes bound.  The Efimov effect is not present for $N>3$.  Still a number of relations and properties exist.  The excited states for each $N$ can appear with mixed spatial symmetry but the ground states are always symmetric.  The first excited state is often also symmetric and these two states are therefore the solutions for identical bosons.  We shall confine ourselves to this symmetry and these two lowest-lying states.

The two lowest states of the tetramer ($N=4$) boson system both lie
below the ground state of the trimer system as seen in
\cref{fig:scat}.  This is in agreement with other numerical
\cite{Platter2004,Hammer2007,Stecher2009,Deltuva2013} and experimental
\cite{Ferlaino2009} investigations.  The spacing strongly indicates
that the second excited state, most likely of mixed spatial symmetry, would appear above the trimer threshold.

Furthermore our results indicate that this behavior continues for $N\geq 4$.  Two pentamer ($N=5$) states accompany the ground state of the tetramer system, and the first excited state of the pentamer system lies so close to the ground state of the tetramer state, that it again seems improbable to find a second excited pentamer state below the tetramer threshold. 

We now add one more particle to study the hexamer ($N=6$) system.  The behavior is precisely repeated and we find two states below the pentamer ground state.  This pattern is also found in a previous work \cite{Gattobigio2012}, where a three-body interaction also is shown to move the first excited states for both $N = 5,6$ above the $(N-1)$-thresholds for binding.  We maintain the use of only two-body interactions throughout this paper. Here it should be mentioned that our results ceases to be universal as the scattering lengths of the system becomes smaller than the range of the potential. The scattering lengths can be seen in \cref{tab:prop} and is further discussed in \cref{sec:uniProp}.

The results for both heptamer ($N=7$) and octamer ($N=8$) systems are also shown in \cref{fig:scat}.  The ground states  systematically appear for decreasing attraction.
However, the first excited states decays into $(N-1)+1$ systems when the scattering length becomes sufficiently small. Thus the excited states follow the $(N-1)$-body ground state, as the last particle is infinitely far away from the $N-1$ particles.
The pattern of two states below the $(N-1)$ binding threshold is broken for $N \geq 7$.  This disruption of an apparently almost universal behavior requires an understanding of the spatial structure.
We shall return with schematic models in the next section but let us
first in the next subsections extract a number of universal quantities.

\begin{table*}
	\centering
	\caption{Properties of the $N$-body systems at the threshold of binding calculated using the stochastic variational method. Column two to four shows the scattering lengths and their relations. Columns six and seven show the average distance between nearest neighboring particles. The eighth column is the relation between the average distance between nearest neighboring particle of the two lowest states. The last two columns give the ratio of the energy at unitarity, compared to the three-body ground state and the $N+1$ excited state.}
	\label{tab:prop}
	\begin{ruledtabular}
		\begin{tabular}{l|cccc|ccc|cc}
			$N\;\;$ & $a_N^{(0)}/b$ & $a_N^{(1)}/b$ & $a_{N}^{(0)}/a_{N-1}^{(0)}$ & $a_{N}^{(1)}/a_{N-1}^{(0)}$ & $\langle r_\text{d}^{(0)}/b \rangle$ & $\langle r_\text{d}^{(1)}/b \rangle$ & $\langle r_\text{d}^{(0)}/r_\text{d}^{(1)} \rangle$ & $E_N^{(0)}/E_3^{(0)}$ & $E_{N}^{(1)}/ E_{N-1}^{(0)}$\\
			\hline
			3 & $-4.395$ & $-92.949$ & - & - & 5.913 & 50.369 & 0.117 & 1 & - \\
			4 & $-2.005$ & $-4.144$ & 0.456 & 0.943 & 2.302 & 5.582 & 0.412 & 5.87 & 1.038 \\
			5 & $-1.283$ & $-1.964$ & 0.635 & 0.980 & 1.790 & 3.349 & 0.535 & 16.01 & 1.092 \\
			6 & $-0.959$ & $-1.263$ & 0.753 & 0.993 & 1.376 & 2.480 & 0.555 & 32.09 & 1.115\\
			7 & $-0.814$ & $-1.065$ & 0.849 & 1.111 & 1.038 & 1.490 & 0.697 & 53.68 & 1.128\\
			8 & $-0.661$ & $-0.980$ & 0.812 & 1.204 & 0.992 & 1.153 & 0.860 & 84.32 & 1.156 \\
		\end{tabular}
	\end{ruledtabular}
\end{table*}

\subsection{Universal quantities}\label{sec:uniProp}

The present finite-range model uses by definition a unit of length, $b$. However if one used another potential with a different unit of length, the absolute results could be superficially different, but the ratios are expected to remain largely the same when universality conditions are approached.
Proper universal numbers are obtained by ratios forming dimensionless quantities. The most direct model-independent parameters are scattering lengths as presented in \cref{tab:prop}.  If universality is a valid concept then for example the ratio of scattering lengths would be true universal number, meaning that other potentials must produce the same ratios.

The threshold value for binding of the dimer is $\pm \infty$ while finite values are obtained for larger particle numbers, $N$. In units of $b$ we see that the sizes of the scattering lengths, $a$, decrease from larger to smaller than $b$ where $N \simeq 6$ and $8$ is the crossing point for ground and excited state, respectively. When $a$ is smaller than $b$ the details of the potential are important and universal properties cannot be expected.  A striking feature in \cref{tab:prop} is that the scattering lengths of the ground state for $N$ and the first excited state for $(N+1)$ are almost equal until $N \geq 7$.

The relations between the trimer and tetramer scattering lengths agree well with the values predicted theoretically in \cite{Stecher2009,Gattobigio2012,Deltuva2013}, and measured experimentally in
\cite{Ferlaino2009}.  The relation between the tetramer and pentamer
scattering lengths is close to the results found numerically in
\cite{Stecher2010,Gattobigio2012}. The deviations are due to the repulsive three-body potential employed in these references, where we in contrast only use an attractive two-body potential. The three-body potential is used in order to simulate a zero-range potential more closely, as it is tuned to reduce finite-range corrections. We have chosen not to include a three-body interaction as we do which to simulate finite-range interactions, which must be closer to the real interactions than a zero-range potential.  The relations are also in agreement with experimental observations \cite{Zenesini2013}.  All the scattering length ratios ($N=3,4,5,6$) also agree with the ones found in \cite{Stecher2011}.

The energies are directly observable and often presented as the results of calculations. Again, ratios are the most model independent quantities as given in \cref{tab:prop} at unitarity.  The ground state binding energies increase rather dramatically with $N$.  Our actual values for $N\leq 6$ are all larger than the results obtained in \cite{Stecher2010,Gattobigio2012}, however the discrepancy can again be attributed the repulsive three-body potential employed there.  The effect of this three-body potential is directly shown in \cite{Gattobigio2012}, and discussed in \cite{Stecher2011}, where $E_6^{(0)}/E_3^{(0)} \sim 30$, as in \cref{tab:prop}, is found with only a two-body potential. We also notice that our results for $N=4$ are larger than the well-known results obtained in \cite{Stecher2009}, this is again due to finite-range correction, and it is worth noticing that these corrections are larger for the more dense ground state, than for the more dilute excited state, where the deviations are less than 3\%. Comparing the results of \cite{Gattobigio2014}, who also employs two-body finite-range Gaussian interactions, we find that our results match exactly for $N=5,6$, while the four-body state is very close to their results. It is striking that the ground state for $N$ and the first excited state for $(N+1)$ have very similar energies as shown in \cref{tab:prop}.  Again these energy ratios are moderately larger than in \cite{Gattobigio2012}.

\subsection{Spatial properties }

\begin{figure}
	\centering
	\includegraphics[width=\columnwidth]{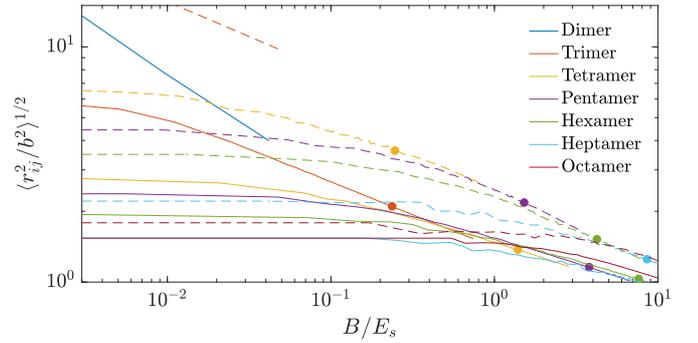}
	\caption{Plot of the root-mean-square distance between the particles in units of $b$, for $N$ identical particles. The solid lines are the ground states, while the dashed lines are the first excited states. The filled circles indicate where the two-body scattering length becomes infinitely large. The noise at higher binding energies is due to numerical imprecisions.}
	\label{fig:idenDist}
\end{figure}

The simplest quantity with information about the structure of the
solution is the second radial moment, that is the mean-square-radius
defined as
\begin{equation}
N \langle r^2 \rangle =
\langle \sum_ {i=1}^{N}(\mathbf{r}_i -  \mathbf{R}_c)^2 \rangle \; ,
\end{equation}
where $\mathbf{r}_i$ and $\mathbf{R}_c$ are coordinates of particle
$i$ and the center-of-mass, respectively.  This average radius is
related to the average distance, $\langle r_{ij}^2 \rangle$, between
particle $i$ and $j$ by the simple expression
\begin{align} \label{eq:dist}
\left\langle r_{ij}^2 \right\rangle = \frac{2N}{N-1} \left\langle r^2 \right\rangle.
\end{align}
It may appear a little strange that the interparticle distance and the root-mean-square radius of the whole system are of similar size. However, here we must remember that the average distance from any particle, $i$, to any other particle, $j$, include contributions from both close-lying and distant configurations.

We show in \cref{fig:idenDist} the interparticle average distance as
function of the binding energy, $B = -E$, for each $N$. The natural
length and energy units are used in the figure.  The dimer
mean-square-radius diverge as $1/B$ as a universal two-body halo state \cite{jen04}. The trimer radius looks like it is finite, however a closer inspection shows that it diverges logarithmically for $B \rightarrow 0$ as a universal three-body halo, which is in agreement with \cite{Fedorov1994}. On the \cref{fig:idenDist} this is most easily seen for the ground state since the radius of the excited state still increases strongly on this scale.

When $N \geq 4$ all radii, both for ground and excited states, increase
smoothly towards a constant when their respective binding energies
approach zero.  In the opposite limit with binding energies around the
natural values, $E_s$, also the radii have a size comparable to the
interaction range, $b$. This is understandable since the entire wave
function then in all cases must be limited by the potential size.  The
radii decrease monotonically with increasing $N$.  The systems become
denser for larger number of particles, and for $N=8$ the radius is
only slightly larger than the range of the interaction.  All pairs
would then be interacting in the ground states for $N \geq 8$.  All
these results are for $N \leq 6$ in agreement with the theoretical
results from \cite{Yamashita2011}.

\begin{figure}
	\centering
	\includegraphics[width=\columnwidth]{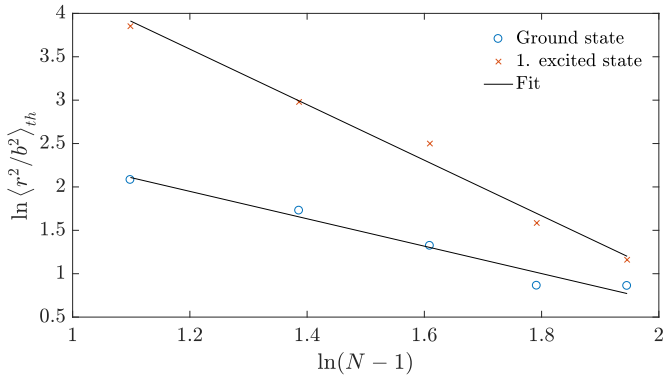}
	\caption{Logarithmic plot of the maximal (at the threshold of binding) root mean-square-radius of the systems as function of the particle number. The solid line is a fit of the type in \cref{eq:powerLaw}. The deviations from the fit is due to numerical imprecisionn.}
	\label{fig:maxR_N}
\end{figure}

The sizes are monotonic and systematic at the binding thresholds where
the energies are zero and the sizes are maximal.  We show the
corresponding radii in \cref{fig:maxR_N} as functions of the number of
particles for both ground and excited states.  This plot is linear on
the logarithmic scale, and the dependence is therefore of the form
\begin{align}\label{eq:powerLaw}
\left\langle r^2/b^2 \right\rangle_\text{th} = C (N-1)^p
\end{align}
where $C$ and $p$ are constants.  A numerical fit leads to
$(p_0,C_0) = (\SI{-1.40 \pm 0.25}{},\SI{14.6 \pm 1.5}{})$ and
$(p_1,C_1) = (\SI{-2.94 \pm 0.09}{},\SI{455.6  \pm 1.2}{})$ for ground and
excited states, respectively.  The uncertainties are due
to the fits, and we can conclude that the numerically extracted radii
are consistent with
\begin{align}\label{eq:powerLaw1}
\left\langle r^2\right\rangle_\text{th}^{1/2} =& \; 3.8 b (N-1)^{-3/4}, \\
\left\langle r^2\right\rangle_\text{th}^{1/2} =& \; 21.3 b (N-1)^{-3/2},
\label{eq:powerLaw2}
\end{align}
for ground and excited states.

\subsection{Assessing universality conditions }

The average radii in \cref{eq:dist,eq:powerLaw1,eq:powerLaw2} are only average measures of distances
between neighboring particles.  Universality requires a substantial
part of the wave function in classically forbidden regions where the
details of the potentials are unimportant. We must therefore estimate
the distances between neighboring particles compared to the range of
the interaction.

A constant density, $\rho_0$, with a total of $N$ particles inside a
sharp cut-off radius, $R$, means $4 \pi R^3 \rho_0 / 3 = N$, where the
mean-square-radius is given by $\langle r^2 \rangle = 3R^2/5$. Inside this
distribution the radius, $r_d$, of a sphere which on average contains
precisely one particle is given by $4 \pi r_d^3 \rho_0 / 3 = 1$. We
then immediately conclude that
\begin{align} \label{eq:raddist}
r_\text{d}^{(n)} = \frac{\sqrt{5/3}}{N^{1/3}}\left\langle r^2 \right\rangle^{1/2},
\end{align} 
where $n=0,1$ labels ground and excited states.

These average distances between nearest neighboring particles at threshold are given in \cref{tab:prop} as functions of $N$ for both ground and excited states. To occupy the classically forbidden regions and ensure universality, we must have that $r_d$ is significantly larger than the interaction range, $b$.  This is true for all systems, but for the heptamer and octamer ground states this condition is severely violated since $r_\text{d}^{(i)} \simeq b$.  The ground states for $N \geq 7$ cannot be universal and therefore not expected to have the same structure with another interaction. This conclusion is consistent with the calculated scattering lengths significantly smaller than $b$ at the threshold of binding. On the other hand, it is remarkable that the universality condition is obeyed for $N \leq 6$ even at the threshold of zero binding energy.

\section{Schematic models} \label{sec:schem}	

Qualitative estimates in schematic models are illustrative.  Two extremes are the short-range Gaussian and the long-range oscillator potentials.  The oscillator potentials can be solved analytically but formally only valid for bound states.  The Gaussian potential is used in the numerical calculations but approximate analytic properties can be derived.  The idea is to the derive properties of the two lowest-lying states in order to indicate whether energies and wave functions of the numerical calculations are universal for a given number of particles.

\subsection{Oscillator estimates}

The effective barrier in the lowest hyperspherical potentials for $N
\geq 4$ tend to give confined bound state structures which allow an
oscillator approximation. The Hamiltonian can be written as
\begin{align*} 
H_\text{osc} =& \sum_{i=1}^{N} \frac{\mathbf{p}_i^2}{2m} - \frac{\mathbf{P}^2}{2m N} \\ 
&+ \frac{1}{4} m \omega^2  \sum_{i<j}^{N} (\mathbf{r}_i -\mathbf{r}_j)^2
- \frac{1}{2}N(N-1)V_N  \\ 
=& \sum_{i=1}^{N} \frac{(\mathbf{p}_i-\mathbf{P}/N)^2}{2m}  \\ 
&+ \frac{1}{4} m \omega^2
\sum_{i=1}^{N} (\mathbf{r}_i -\mathbf{R}_c)^2
- \frac{1}{2}N(N-1)V_N \, ,
\end{align*}
where $V_N$ is the energy shift of each interacting pair, and the
reduced mass is $m/2$.  The center-of-mass coordinate, $\mathbf{R}_c$,
and total momentum, $\mathbf{P}$, are defined by
\begin{align}
\sum_{i=1}^{N} \mathbf{r}_i  =  N  \mathbf{R}_c \, , \qquad
\sum_{i=1}^{N} \mathbf{p}_i  =  \mathbf{P} \; .
\end{align}
The energies, $E_N^{(n)}$, and mean-square-radii for the $n$th
$N$-body system are then
\begin{align}
E_N^{(n)} =& \; \hbar \omega \sqrt{\frac{N}{2}} \left(n+\frac{3}{2}\right) N - \frac{1}{2}N(N-1)V_N \, , \\
&\left\langle(\mathbf{r}_i- \mathbf{R}_c)^2\right\rangle = \frac{\hbar}{m \omega} \sqrt{\frac{2}{N}} \left(n+\frac{3}{2}\right)\,.
\end{align}
Thus the frequency for the one-body motion in the $N$-body system is
multiplied by $\sqrt{N/2}$. This implies that the mean-square-radii
are divided by this quantity. The general radius decrease is by the oscillator estimate to be $N^{-1/4}$.

Furthermore, the energies of the two lowest states for the $N$-body
system are
\begin{align}\label{eq:E0osc}
E_{N}^{(0)} =& \;\hbar \omega \frac{3N}{2}\sqrt{\frac{N}{2}} - \frac{1}{2}N(N-1)V_N \, , \\ \label{eq:E1osc}
E_{N}^{(1)} =& \;\hbar \omega \frac{5N}{2}\sqrt{\frac{N}{2}} - \frac{1}{2}N(N-1)V_N \, . 
\end{align}
Then clearly the threshold where the ground state energy is zero,
$E_{N}^{(0)}=0$, requires that the strength, which plays the role of a
potential strength, is
\begin{align}\label{eq:VNth}
V_{N}^\text{th} = \hbar \omega \frac{3}{N-1} \sqrt{\frac{N}{2}} \; .
\end{align}
The system becomes more bound with larger $V_N$ or smaller $\omega$, where $\omega$ entirely determines all radii. The question in the present context is where the excited states of the $(N+1)$-body system appear with a strength corresponding to the threshold of the $N$-body system.

Inserting the potential at threshold, \cref{eq:VNth}, in the above expressions for the energy of the two lowest states, \cref{eq:E0osc,eq:E1osc} we find
\begin{align} 
&\begin{aligned}
E_{N+1}^{(0)}(V_{N}^\text{th})  =&\; \hbar \omega   \frac{3(N+1)}{2} \sqrt{\frac{N}{2}} \\
&\times \bigg( \sqrt{\frac{N+1}{N}} - \frac{N}{N-1} \bigg) \, ,
\end{aligned} \\ &
\begin{aligned}
E_{N+1}^{(1)}(V_{N}^\text{th})  =&\; \hbar \omega   \frac{3(N+1)}{2} \sqrt{\frac{N}{2}} \\
&\times \bigg(\frac{5}{3} \sqrt{\frac{N+1}{N}} - \frac{N}{N-1} \bigg) \, .
\end{aligned}
\end{align}
Expanding the expressions in the bracket for large $N$, we find
$-1/(2N)$ and $1/(6N)$, respectively. Thus the first excited state is
one third of the ground state energy away from zero on the unbound
side, that is relatively close to zero. The infinite oscillator walls
push the state upwards. If the strength had been increased to $V_{N}
\simeq V_{N}^\text{th}(1+1/(6N))$, the excited state would have been
at zero energy. Now it is slightly unbound.

\begin{table*}
	\centering
	\caption{Comparison of various quantities of the $N$-body state with the $(N+1)$-body halo state at threshold calculated using the results from the stochastic variational calculation. The second column shows the potential strength at threshold of binding for the $N$-body system. The third and forth column shows the binding energy of the $(N+1)$-body states at threshold. The fifth and sixth column shows the squared radius of the systems at threshold. The seventh column shows the halo radius calculated using \cref{eq:rhalo}. The eight column is the binding energy of the halo system calculated using \cref{eq:bindHalo}. Column nine shows $G_\text{crit}$ of the Gaussian effective potential, as found in the left side og the approximation of \cref{eq:constrain}, while the last column displays the halo scattering length of the $(N+1)$-body system. Energies and lengths are in units of $E_s$ and $b$ respectively.}
	\label{tab:halo}
	\begin{ruledtabular}
		\begin{tabular}{l|ccccccccc}
			$N$ &  $V^{(\text{thres})}_0$ & $B_{N+1}^{(0)}$ & $B^{(1)}_{N+1}$ & $\left\langle r^2\right\rangle^{(0)} _N$ & $\left\langle r^2\right\rangle^{(1)}_{N+1}$ & $\left\langle r^2\right\rangle_\text{halo}$ & $B_\text{halo}$ & $G_\text{crit}$ & $a_\text{halo}$\\
			\hline
			2 & $\SI{-1.353(9)}{}$ & 0.253 & $\SI{9e-4}{}$ & $\SI{8.71e2}{}$ & $\SI{3.16e3}{}$ & $\SI{7.76e3}{}$ & $\SI{4.8e-5}{}$ & $0.075$ & $-11.1$\\
			3 & $\SI{-1.066(7)}{}$ & 0.456 & $\SI{5.3e-3}{}$ & 43.6 & 47.1 & 57.5 & $\SI{5.8e-3}{}$ & $0.437$ & $-3.0$\\
			4 & $\SI{-0.853(8)}{}$ & 0.567 & $\SI{8.6e-3}{}$ & 8.0 & 19.7 & 66.3 & $\SI{4.7e-3}{}$ & $1.085$ & $-12.0$\\
			5 & $\SI{-0.704(7)}{}$ & 0.582 & $\SI{7.1e-3}{}$ & 5.6 & 12.2 & 45.0 & $\SI{6.7e-3}{}$ & $1.345$ & $1172.6$\\
			6 & $\SI{-0.607(7)}{}$ & 0.531 & $-0.245$ & 3.7 & 4.9 & 11.6 & $\SI{2.5e-2}{}$ & $1.669$ & $10.5$\\
			7 & $\SI{-0.553(5)}{}$ & 1.014 & $-0.951$ & 2.4 & 3.2 & 9.0 & $\SI{2.5e-2}{}$ & $2.109$ & $4.7$\\
		\end{tabular}
	\end{ruledtabular}
\end{table*}

\subsection{Gauss density distribution estimate}

We assume a Gaussian density distribution of $N$ particles, with a range of $r_G$ and a strength of $\rho_0$
\begin{align} \label{eq:gaussdist}
\rho (r) = \rho_0 e^{-r^2/r_G^2},
\end{align}
with the requirement
\begin{align}
\int \rho(r) d\tau = N,
\end{align}
where $\tau$ denotes the integration over all coordinates.
The mean-square-radius, $R_N$, of the distribution in
\cref{eq:gaussdist} is related to $r_G$ by
\begin{align}\label{eq:RN2}
R_N^2 = \frac{\int \rho(r) r^2 d\tau}{\int \rho(r) d\tau}
= \frac{3}{2}r_G^2.
\end{align}

We assume the first excited state consists of a core of $N$ particles
plus a very weakly bound additional particle.  In total forming a
two-body halo state of very small binding energy.  We fold the
Gaussian potential of the last particle and the Gaussian density
distribution in order to obtain the effective  potential, that
is
\begin{equation}\label{eq:Veff}
\begin{aligned}
V_\text{eff}(r) =& V_0\rho_0 \int e^{-(r-r')^2/b^2}e^{-r'^2/r_G^2}d\tau' \\
=& \frac{NV_0b^3}{(b^2 + r_G^2)^{3/2}}
\exp\left(- \frac{r^2}{b^2 + r_G^2}\right).
\end{aligned}
\end{equation}
If this Gaussian effective potential should have infinite scattering length the parameters must be constrained by
\begin{align}\label{eq:constrain}
G_\text{crit} \equiv\frac{\mu_N N V_0 b^3}{\hbar^2 (b^2+ 2R_N^2/3)^{1/2}}  \simeq 1.34,
\end{align}
where $1.34$ corresponds to the threshold of binding for a two
particle system interacting with Gaussians, and $\mu_N = mN/(N+1)$ is
the reduced mass of the halo two-body state.

All quantities on the left hand side of \cref{eq:constrain} are known and the validity of the relation can be tested. Furthermore, the combination of these quantities determine uniquely the scattering length for a Gaussian interaction in units of its range, $(b^2+ 2R_N^2/3)^{1/2}$. This is an indirect and perhaps very inaccurate and impractical way of finding the scattering length between one particle and the system of $N$ particle at threshold.

A number of characteristic results obtained using the results of the stochastic variational calculations are given in \cref{tab:halo}. The second column shows the potential strength at the threshold of binding. The third and forth columns display the binding energies of the $(N+1)$-body state at threshold. This is found as the energy of the $(N+1)$-body system at the threshold of binding for the $<N$-body system. The fifth and sixth columns show the squared radii of the systems at threshold.

The structure of an excited state weakly bound to an almost unperturbed core of $N$ particles can be tested directly by comparing binding energy and radius of this halo structure.  The halo radius, $\left\langle r^2 \right\rangle_\text{halo}$, is found from $R_N$ and the total radius of the excited $(N+1)$-body by \cite{jen04} 
\begin{align}\label{eq:rhalo}
\left\langle r^2  \right\rangle^{(1)}_{N+1} = \left\langle r^2
\right\rangle^{(0)}_{N}
\frac{N m}{N m + M}
+ \left\langle r^2 \right\rangle_\text{halo} \frac{M}{Nm + M}.
\end{align}
This radius is calculated by use of the stochastic variational results, and shown in column seven of \cref{tab:halo}. This halo radius is inversely proportional to the binding energy, $B_\text{halo}$, of the halo state \cite{jen04}, that is
\begin{align}\label{eq:bindHalo}
\left\langle r^2 \right\rangle_\text{halo} =
\frac{\hbar^2}{4 \mu_N B_\text{halo}}.
\end{align}
Thus the structure can be tested by deriving the binding energy from the radius using \cref{eq:bindHalo} and \cref{eq:rhalo}, and comparing to the directly calculated values. This yields the values in column eight of \cref{tab:halo}.

The ninth column of \cref{tab:halo} shows the value of $G_\text{crit}$ calculated using \cref{eq:constrain}. This value varies by a factor of about $2$ around the critical value of $1.34$. This $N$-dependence originate from $NV_0$ and the mean square radius.  Looking closer into the values in \cref{tab:halo} we notice that $NV_0$ is only marginally increasing with $N$ while the rather strong decrease of $R_N$ therefore must be responsible for the increase of $G_\text{crit}$.  The uniquely related scattering lengths in \cref{tab:halo} show the variation through the divergence from finite negative to positive values.

The increase of $G_\text{crit}$ shows increasingly stronger one-body binding with $N$.  The corresponding remarkable change through $1.34$ from unbound to bound one-body states is precisely opposite to the numerical results.  This indicates that the one-body picture and the corresponding halo structures only are coincidental passing properties which quickly disappear with increasing $N$. This is also in line with the behavior of the energy of the first excited state for $N>6$.

The fact that $G_\text{crit}$ varies over a relatively large interval simply reflects that fact that the Gaussian effective potential does not have an infinite scattering length. The last column in \cref{tab:halo} shows the corresponding halo scattering length of a $(N+1)$-body i.e. the scattering length between the closely bound $N$-body system and the loosely bound last particle. This is found by modelling the corresponding $G_\text{crit}$ at the two-body potential strength for the halo state, and then calculating the corresponding halo scattering length. For $N<5$ the halo scattering length becomes negative and hence unbound, while it becomes bound for larger $N$. These numbers are rather small for all $N$ with the exception of $N=5$, where the halo scattering length is significantly larger than for the rest. This could also have been expected looking at $G_\text{crit} = 1.345$, which is close to the analytical value of 1.34.

The results collected in \cref{tab:halo} show that all radii decrease
with $N$.  The ground state energy varies relatively little for these
threshold strengths of the $N$-body system.  Furthermore, the binding
energy for the excited state is very close to zero for all $N \leq 6$.
The predicted halo bindings for $N=3,4,5,6$ are again all rather small
and all within a factor of two from the true directly calculated
values.  In fact, this comparison is even more favorable, since the
uncertainty on the strength (given in second column of
\cref{tab:halo}) reflects an uncertainty of the same order as the
discrepancies in \cref{tab:halo}.

The halo radii in \cref{tab:halo} decrease substantially with $N$. However, they are all large compared to the interaction range and therefore exhibiting characteristic halo structure.

\section{Mass variation of one particle}\label{sec:mass1}

The symmetric structure is necessary for identical bosons. Another degree of freedom is introduced by allowing one particle to be distinguishable. The defining parameters are the mass ratio, $M/m$, between the distinguishable and identical particles, and the two types of pairwise interaction.  To limit the amount of calculations, while still investigating pertinent features, we assume the same potential strength between all particles. We consider only one very large heavy-light scattering length, numerically $|a_{HL}|/b \sim 10^{10}$, and consequently for the same Gaussian we have $a_{HH} = a_{HL} (1+M/m)/2$, which has the same order of magnitude. In other words we are at the unitarity point of zero dimer binding. As we vary the mass ratio $M/m$ to a lower value $a_{HH}$ becomes smaller than $a_{HL}$, however only up to a factor one half for the most extreme case. This also means that the heavy-heavy systems should become more bound, as we would also expect from the fact that the binding energy depends inversely on the reduced mass of the system.

We vary the mass ratio, $M/m$, from $1$ to $1/20$, which is relevant for the Efimov experiments with alkali atoms \cite{Barontini2009,Pires2014,Tung2014}. The opposite case of one heavy and many identical light particles is less interesting since the structure quickly approaches identical particles in a dominating one-body field.  The number of identical particles is $N \leq 7$ for practical calculational reasons.

\begin{figure}
	\centering
	\includegraphics[width=\columnwidth]{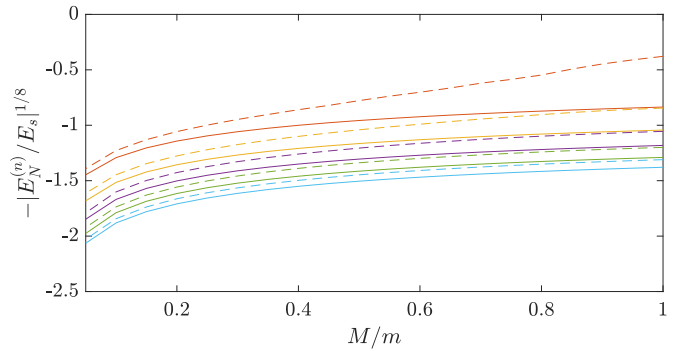}
	\caption{Plot of the energy the heteroparticle systems at unitarity, in units of the characteristic energy. As for the identical particles in \cref{fig:scat} there is no crossing between the different systems, when the mass is changed. Notice that the characteristic energy $E_s$, from \cref{eq:CharE}, uses the identical two-particle reduced mass, such that the evolution of the energies is not hidden behind a change in the characteristic energy. The color scheme in this figure is the same as in \cref{fig:scat}.}
	\label{fig:Energy}
\end{figure}

\begin{figure}
	\centering
	\includegraphics[width=\columnwidth]{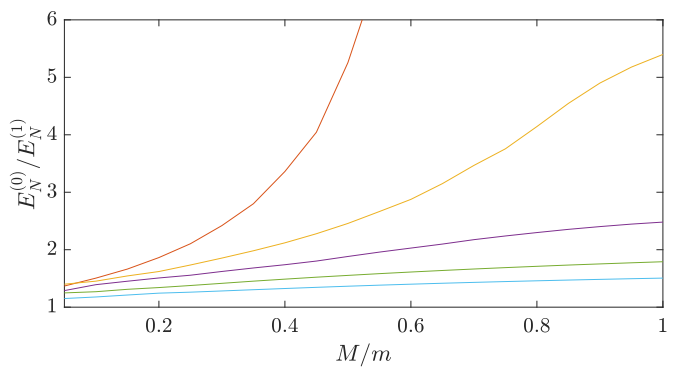}
	\caption{Plot of the energy ratio between the ground state and the first excited state of the heteroparticle systems. The color scheme in this figure is the same as in \cref{fig:scat}.}
	\label{fig:EnergyRatio1}
\end{figure}

\subsection{Energies}

The most important quantity is the energy. Theoretical studies suggest that the two tetramer states accompanying the identical ground state trimer are also attached for different mass ratios \cite{Blume2014,Schmickler2017}. We extend these investigations to larger $N$. The resulting energies are shown in \cref{fig:Energy} for $N$-body systems as functions of mass ratio.  Decreasing the mass ratio from unity, all energies decrease implying that the systems become more bound. As with the identical particles, the ground and first excited states remain together below the ground state energy of the system with one less particle. We also notice that the distance between the two states decreases, while the distance increases to the accompanying state.

In \cref{fig:EnergyRatio1} we show the energy ratio, $E_N^{(0)}/E_N^{(1)}$, of the ground and first excited states for $N\leq 7$.  This is in other words the scaling factor between these states. The values are all moderate compared to the trimer scaling for equal masses. However, this is a very misleading comparison, since the structure for the lowest-lying states is incompatible with the trimer states. The concept of scaling factor is therefore very misplaced in the present context. In any case, the ratios decrease
both with decreasing mass ratio \cite{Schmickler2017,Blume2014}, and
with increasing total number of particles. The ratio increases slower
towards unit mass ratio for increasing $N$.

\begin{figure}
	\centering
	\includegraphics[width=\columnwidth]{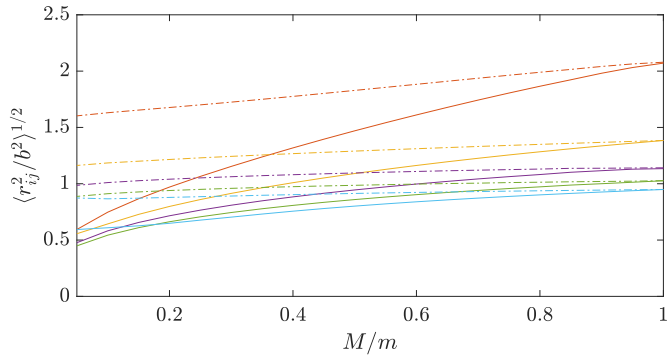}
	\caption{The root-mean-square of the distances,
		$\mathbf{r}_{ij} \equiv \mathbf{r}_i- \mathbf{r}_j$, between
		the particles in the ground states of $(N+1)$-body systems as functions of the mass
		ratio. The solid lines are the identical distances, while
		the dashed-dotted lines are the interspecies distances. The color scheme in this figure is the same as in \cref{fig:scat}.}
	\label{fig:massScan_ground}
\end{figure}

\begin{figure}
	\centering
	\includegraphics[width=\columnwidth]{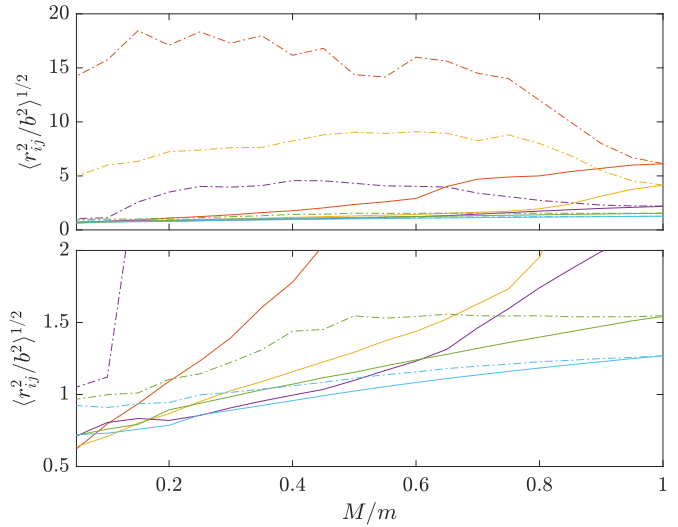}
	\caption{The root mean square of the distances,          $\mathbf{r}_{ij} \equiv \mathbf{r}_i- \mathbf{r}_j$, between    the particles first excited state as a function of the mass ratio. The solid lines are the identical distances, while the dashed-dotted lines are the interspecies distances. The color scheme in this figure is the same as in \cref{fig:scat}. The upper panel shows the full plot, while the lower panel is a zoom of the more dense states. The lack of smoothness on the lines is due to stochastic uncertainties in the numerical algorithm.}
	\label{fig:massScan_1exState}
\end{figure}

\subsection{Radii}

The distances between pairs of particles may be different for
identical and unequal pairs. For the system of $(N+1)$ particles, we
obtain two groups with a degeneracy of $N$ and $N(N-1)/2$,
respectively for unequal and equal particle pairs. We show in
\cref{fig:massScan_ground} the interparticle distances of the ground
states for $(N+1)$ particles as functions of mass ratio.

The two groups of distances coincide for equal masses, and all
distances decrease for decreasing mass ratio. However, distances
between unequal pairs decrease much less than distances between equal
pairs. This means that the lighter the particle the further it moves
away from the symmetric system of $N$ particles. Or more correctly,
the symmetric boson system becomes denser while the light particle
essentially remains where it was for equal masses.

The sizes of the symmetric subsystems approach each other as the mass
ratio decreases, except for the $N=6+1$ system which seems to saturate at
a constant size for mass ratios below $0.2$. This is consistent with
a structure where all these particles are located inside the range of
the two-body potential, and therefore only confined spatially by the
walls of this potential. The light particle moves slower towards
smaller distances, and perhaps even towards a similar size for
different $N$. The structure begins to resemble one dense core
surrounded by the light particle at a distance of at least twice the
core radius.

The pair distances for the first excited states are shown in
\cref{fig:massScan_1exState}. Again all pair distances coincide for
equal masses. The distance between equal pairs systematically decrease
smoothly with mass ratio, similar to the behavior of the ground
state. The light particle behaves apparently very individual depending
on the number, $N-1$, of the identical particles. We now find that
distances between unequal pairs even increase for decreasing mass
ratio.  This is especially pronounced for $N=4$ and $5$ where a rather
large and relatively slowly varying value is obtained over a large
mass ratio interval. However, for $N=4$ this distance suddenly drops
for a mass ratio below $0.1$.

For $N=6$ and $7$ the increase for mass ratios close to unity is only
marginal or absent. Furthermore, the light particle distances fall
below the distances between the identical particles almost immediately
for $N=7$ and below a mass ratios of $0.4$ for $N=6$. This means that
the structure changes completely from a few dilute identical particles
surrounded by a light particle far away to a light particle moving
within the dense core of identical particles.

\begin{figure}
	\centering
	\includegraphics[width=\columnwidth]{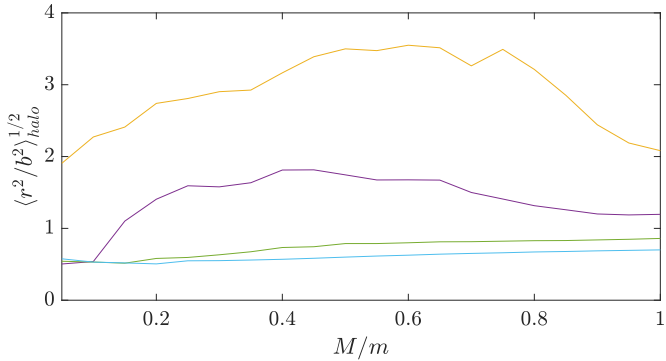}
	\caption{The halo radii, found using \cref{eq:rhalo}, at the unitarity limit as function of mass ratio, $M/m$. The color scheme in this figure is the same as in \cref{fig:scat}. The lack of smoothness is due to stochastic uncertainties.}
	\label{fig:rHalo}
\end{figure}

\subsection{Structure of the excited states}

The structure of the excited states could perhaps be two-body halo
states even at the unitarity limit.  An efficient and simple test
of this is to compare fully numerically calculated binding energies,
$B_{N+1}^{(1)} - B_{N}^{(0)}$, with those, $B_\text{halo}$, obtained
from \cref{eq:bindHalo}, where the halo radius is found from
\cref{eq:rhalo} and \cref{eq:dist}.  This was already done in
\cref{tab:halo} at the thresholds of zero $N$-body binding, that is far away from the interaction corresponding to zero dimer binding energy.

First the anticipated halo radii found from \cref{eq:rhalo} are shown
in \cref{fig:rHalo}.  The behavior changes both qualitatively and
quantitatively from $N=4,5$ to the higher values of $N$. The
root-mean-square radii for $N=4,5$ are $2-3$ times larger than the
interaction range and these states can be classified as halo
structures. On the other hand, all these systems for $N>5$ are smaller
than the core.  This is in contrast to the structure at the thresholds for
binding of the $N$ core particles.

\begin{figure}
	\centering
	\includegraphics[width=\columnwidth]{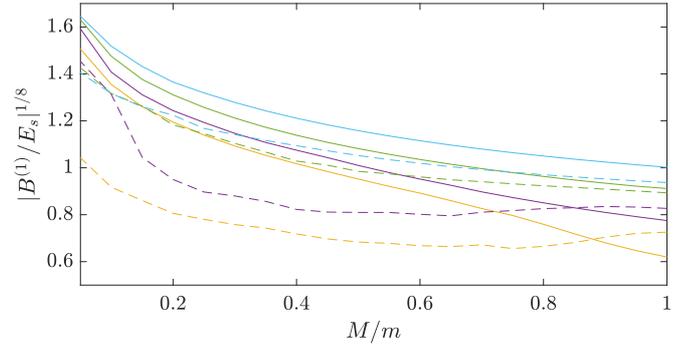}
	\caption{The binding energies of the first excited states as
		functions of mass ratio at unitarity limit. The solid curves are $B_{N+1}^{(1)} - B_{N}^{(0)}$ and the dashed
		curves are obtained from \cref{eq:rhalo,eq:bindHalo}. The color scheme in this figure is the same as in \cref{fig:scat}.}
	\label{fig:haloenergy}
\end{figure}

The corresponding binding energies are shown as functions of mass
ratio in \cref{fig:haloenergy}. We emphasize this figure is for
interactions at unitarity limit, that is for infinite two-body scattering
length.  The overall property is that the schematic energies obtained
from the halo-radius assumption are systematically about $20\%$ below
the correctly calculated binding energies.  The more versatile lowest
$N=4,5$ results corresponding to halo structure deviate by up to twice
as much.  There is no systematic tendency towards better agreement for
any larger or smaller values of the mass ratio.  These results at
unitarity limit are for $M/m=1$ consistent with those obtained for zero
binding energy.

\begin{figure}
	\centering
	\includegraphics[width=\columnwidth]{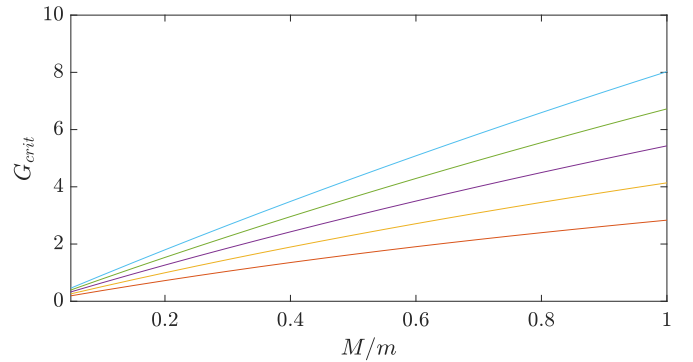}
	\caption{The critical number, $G_\text{crit}$, at the unitarity limit as function of mass ratio, $M/m$. The colors of the lines represent the same systems as in
		\cref{fig:scat}.}
	\label{fig:gcrit}
\end{figure}

Another revealing quantity is $G_\text{crit}$ in \cref{eq:constrain}, where
$G_\text{crit} \simeq 1.34$ would imply an infinitely large scattering length
and hence zero binding energy.  In the unitarity limit where the
two-body system has a state of zero energy the parameters are
constrained by
\begin{align}\label{eq:constrain1}
\frac{m M |V_0| b^2}{\hbar^2 (M+m)}  \simeq 1.34,
\end{align}
which combined with \cref{eq:constrain} for the $N$-body system
leads to the expression
\begin{equation}\label{eq:constrain2}
\begin{aligned}
& G_\text{crit}(N,M/m) =\\
& \frac{ 1.34 (1+ M/m)}{(1+M/(m(N-1))) (1 + 2(R_{N-1}/b)^2/3)^{1/2}} \; .
\end{aligned}
\end{equation}
This result is shown in figure \cref{fig:gcrit} as function of both
$N$ and $M/m$.  The explicit $N$ and reduced mass dependencies explain
most of the increases with both $N$ and for $M \rightarrow m$. The only
other quantity is the mean-square-radius of the $N$-body system which
decreases and consequently enhances the explicitly increasing factors.

The increase with $N$ to far above $1.34$ strongly suggests that at
least the energies must change. For a square well, $9$ times the
critical number (which is $1.34$ for a Gaussian) would correspond to
the next infinitely large scattering length.  For $N=6$ and $M=m$ we
are already halfway to this value where we know that the radius is
comparable to the core size, and the state becomes unbound.  This
means that the scattering length corresponding to the unequal particle
colliding with the ground state structure of the $N$ particles quickly
becomes finite and approaching the range of the two-body potential.

It is interesting to note  that for $m \simeq 20 M$, we find values of $G_\text{crit}$ in the neighborhood of the critical value $1.34$.
This indicates halo structure for these states. In the other end of equal masses, $m \simeq M$, we might reach the next infinite
scattering length corresponding to $G_\text{crit} \simeq 9 \times 1.34$, that is again a possible halo structure for these states.

\section{Conclusion}

We have calculated the two lowest-lying weakly bound states of systems of $N$ identical bosons, where $N$ is between 3 and 8, for different scattering lengths.  We use a finite-range Gaussian potential and search for universal structures and limits to applicability of this concept. The range of the potential is used as the unit of length, and the strengths decrease from values corresponding to infinite scattering length to threshold of binding.

The calculated ratios between ground and first excited state energies at the unitarity limit are, due to the finite range, slightly larger than the predicted Efimov scaling factor.  The same effect is still present at the thresholds where it is seen in the calculated scattering lengths, also deviating  from Efimov scaling predictions. The systematics are in agreement with previous works in the literature for $N \leq 6$.

We find that two $(N+1)$-states accompany each ground state for all $N$ at the unitarity limit.  When the two-body attraction decreases the bound states move into the continuum one by one while $N$ is increasing from $3$. Still both ground and first excited states are found below the ground state of the $N$-body system, until the systematics is broken for $N \geq 6$. The first excited $(N+1)$-body state exceeds the $N$-body ground state when $N \geq 6$.

Information about the structure of these quantum states can be found
in their radii, which monotonically increase with decreasing
attraction as the threshold for binding is approached.  We find the
well established strong divergence of the two-body system and the
logarithmic divergence of both states for $N=3$.  In contrast, the
radii of the systems approaching zero binding are found to converge
towards finite sizes for both the two lowest-lying states for all $N
\geq 4$.

The mean-square-radii of ground and excited states at threshold
follow different power law dependences with powers $-3/2$ and $-3$, at
least up to $N=7$.  The structures of the excited states are
consistent with two-body halo structures of one particle interacting
weakly with the dense $N$-body core.  These universal features are
broken for $N \geq 7$ as signaled by a radius smaller than the
interaction range.

The energies and radii are also calculated for systems where one
particle mass is decreased, but only for one interaction with very
large scattering length.  We find for the ground states that the
binding energies increase while the spatial distances decrease as the
mass ratio decrease.  The identical particles form denser
core-structures surrounded by the light particle.

The structures of the first excited states change considerably with decreasing mass ratio. For few particles the core of identical particles is surrounded by the light particle far away. For more than $5$ core-particles the light particle is pulled inside the core
configuration.  These rather peculiar
structure variations are due to the relatively strong attraction
between all particles in the present calculation. This means in
particular that the light particle is subject to a large total
attraction from many identical particles.  The two-body halo structure is only present
for $N \leq 5$.

In summary, we have found several expected structures, a few a
little surprising, some perhaps even strange. We studied identical
boson systems from pairwise bound to the thresholds of $N$-body
binding.  There is clearly a change away from universal structures
when the number of particles exceeds $6$.  We studied a system with
one of the masses decreasing but only with a relatively strong
attraction between all particles.  The ground state structures change
smoothly, while the first excited state structures change rather
dramatically with the number of identical particles.  This is expected
to be different for weaker attraction which is beyond the scope of the
present investigation.

\section*{Acknowledgements}
The authors would like to thank N.~T. Zinner for help in devising 
the project and continuing discussions during its completion
as well as a critical reading of the manuscript. 
This work is supported by the Danish Council for Independent
Research.


\end{document}